\documentclass[12pt,12pt]{article}
\usepackage{graphicx}
\usepackage{epsfig}
\usepackage{amsmath,amssymb}
\usepackage{slashed} 
\catcode`\@=11
\newdimen\ex@
\font\dozeb=cmmib10 scaled \magstep1
\font\dozesyb=cmbsy10 scaled \magstep1
\font\dezb=cmmib10
\def\bm{\fam9}
\font\mathbf cmbxti10 at 12pt

\def\beq{\begin{equation}}
\def\eeq{\end{equation}}
\def\beqa{\begin{eqnarray}}
\def\eeqa{\end{eqnarray}}
\newcommand{\ba}{\begin{eqnarray}}
\newcommand{\ea}{\end{eqnarray}}
\newcommand\BA{\begin{array}}
\newcommand\EA{\end{array}}

\def\vg{{\bm  g}}
\def\vk{{\bm  k}}

\def\vp{{\bm  p}}
\def\vq{{\bm  q}}

\def\vv{{\bm v}}

\def\vx{{\bm x}}

\def\T{{\bm T}}

\begin{document}

\title{\Large {\bf Velocity operator approach to a Fermion system}
\vspace{-0.4cm}}
\author{Seiya NISHIYAMA\footnotemark[1]~
~and
Jo\~{a}o da PROVID\^{E}NCIA\footnotemark[2]\\
\\[-16pt]
Centro de F\'\i sica,
Departamento de F\'\i sica,\\
\\[-14pt]
Universidade de Coimbra,
P-3004-516 Coimbra, Portugal\footnotemark[2]}

\def\bm#1{\mbox{\boldmath $#1$}}
\def\bra#1{\langle #1 |}
\def\ket#1{| #1 \rangle }

\maketitle

\vspace{-1.0cm}

\footnotetext[1]{~$\!$Corresponding author. 

~~ E-mail address: 
seikoceu@khe.biglobe.ne.jp; nisiyama@teor.fis.uc.pt}
\footnotetext[2]{
$\!$  E-mail address:
providencia@teor.fis.uc.pt}

\vspace{0.2cm}


\begin{abstract}
$\!\!\!\!\!\!\!\!\!$In this paper,
we formulate a velocity operator approach
to a three-dimensional (3D) Fermion system.
Following Sunakawa,
introducing density
and velocity operators,
we treat 3D quantum fluid dynamics
in the system.
We get a collective Hamiltonian
in terms of collective variables.
The lowest order collective Hamiltonian
is diagonalized.
This diagonalization leads us to a Bogoliubov transformation
for Boson-like operators.

\end{abstract}

\vspace{-0.2cm}

{\it Keywords}:
 Collective motion of a three-dimensional Fermion system;
 
velocity operator; vortex motion; 
Grassmann numbers

\vspace{0.1cm}

PACS Number (s):
21.60.-n, 21.60.Ev



\def\thesection{\arabic{section}}
\section{Introduction}

\vspace{-0.3cm}

~~
To approach elementary excitations in a Fermi system,
Tomonaga
and Emery
gave basic ideas in their collective motion theories
\cite{Tomo.50,Emery.79}.
On the other hand,
Sunakawa's discrete integral equation method
for a Fermi system
\cite{SYN.62}
may be expected to also work well for a collective motion problem.
In the preceding papers
\cite{NishProvi.15,NishProvi2.16},
introducing
density operators $\rho_{\vk}$ and
associated variables $\pi_{\vk}$
and
defining $exact$ momenta
$\Pi_{\vk}$
(collective variables),
we could get an $exact$ canonically conjugate momenta approach
to one- and three-dimensional (1D and 3D) Fermion systems.
In the present paper,
we formulate a velocity operator approach
to a 3D Fermion system.
Following Sunakawa,
after introducing momentum density operators
$\vg_{\vk}$,
we define velocity operators
$\vv_{\vk}$
which denote classical fluid velocities.
We derive a collective 
Hamiltonian in terms of the collective variables
$\vv_{\vk}$ and $\rho_{\vk}$
for the elementary excitations including vortex motions.
The lowest order collective Hamiltonian is diagonalized.
This diagonalization leads to a Bogoliubov transformation
for Boson-like operators 
\cite{Bogoliubov.47}.

In \S 2
first we introduce collective variables $\rho_{\vk}$ 
and associated momentum density variables $\vg_{\vk}$ and
give the commutation relations between them. 
The velocity operator
$\vv_{\vk}$ is defined by a discrete integral equation
and 
commutation relations between the velocity operators are also given. 
In \S 3
the dependence of the original Hamiltonian on
$\vv_{\vk}$ and $\rho_{\vk}$ is determined.
This section is also devoted to a calculation of a constant term
in the collective Hamiltonian.
Its lowest order is diagonalized and
leads to a Bogoliubov transformation
for Boson-like operators.
Finally in \S 4
some discussions and further outlook are given.


\newpage

\def\thesection{\arabic{section}}
\setcounter{equation}{0}
\renewcommand{\theequation}{\arabic{section}.\arabic{equation}}     
\section{Collective variable and velocity operator}

\vspace{-0.2cm}

~~~~Let us define the Fourier component of the density operator
$(\rho(\vx) \!=\! \psi^{\dag}(\vx) \psi(\vx))$ as\\[-24pt]
\ba
\BA{c}
\rho_{\vk}
\!\equiv\!
{\displaystyle \frac{1}{\sqrt N}} \!
\sum_{\vp} \!
a^{\dag}_{\vp+\frac{\vk}{2}} 
a_{\vp-\frac{\vk}{2}},~
\rho_{0}
\!=\!
{\displaystyle \frac{1}{\sqrt{N}}} \!
\sum_{\vp} 
a^{\dag}_{\vp} a_{\vp} 
\!=\!
\sqrt N ,~
(N\!:\mbox{total number of Fermion}) .
\EA 
\label{FcomponentDensityOp}
\ea\\[-18pt]
We here consider a spin-less Fermion
and have used the anti commutation relations (CR)s among
$a_{\vk}$'s and $a^{\dag}_{\vk}$'s.
The Hamiltonian $H$
in a cubic box  of volume $\Omega~(\!=\! L^3)$
is given as\\[-22pt]
\ba
\BA{c}
H
\!=\!
T \!+\! V
\!=\!
\sum_{\vk} \!
{\displaystyle \frac{\hbar^{2}k^{2}}{2m}}
a^{\dag}_{\vk }
a_{\vk }
\!\!+\!
{\displaystyle \frac{N}{2\Omega}} \!
\sum_{\vk \ne 0}
\nu (\vk)
\rho_{\vk}
\rho_{-\vk}
\!-\!
{\displaystyle \frac{N}{2\Omega}} \!
\sum_{\vk \ne 0}
\nu (\vk)
\!+\!
{\displaystyle \frac{N(N\!-\!1)}{2\Omega}} \!
\nu (0)  .
\EA
\label{ExpressionforHamiltonian}
\ea\\[-18pt]
The Fourier component of the interaction $V$
is given by the relation
$
V \! (\vx)
\!\equiv\!
{\displaystyle \frac{1}{\sqrt{\!\Omega}}} \!
\sum_{\vk} 
\nu (\vk) 
e^{i\vk \cdot \vx}
\!$.

Following Sunakawa
\cite{SYN2.62}, 
we introduce a momentum density operator given by\\[-24pt]
\ba
\BA{c}
\vg_{\vk}
\!\!\equiv\!\!
{\displaystyle \frac{\hbar}{\sqrt N}} \!
\sum_{\vp} \!
\vp 
a^{\dag}_{\vp-\frac{\vk}{2}} \!
a_{\vp+\frac{\vk}{2}}, ~~
\vg_{0}
\!\!=\!
{\displaystyle \frac{\hbar}{\sqrt{N}}} \!
\sum_{\vp}
\vp
a^{\dag}_{\vp} a_{\vp}
\!=\!
0.
\EA
\label{FcomponentMomentumDensityOp}
\ea\\[-16pt]
CRs among $\rho_{\vk}$  
and vector $\vg_{\vk}
(=\! (g_{\vk}^{(1)},g_{\vk}^{(2)},g_{\vk}^{(3)}))$
with vector ${\vk} (=\! (k_1,k_2,k_3))$
are obtained as\\[-22pt]
\ba
\BA{l}
[\vg_{\vk} ,\rho_{\vk'}]
\!=\!
{\displaystyle \frac{\hbar \vk'}{\sqrt{N}}}
\rho_{\vk'-\vk} ,~~
[g^{(i)}_{\vk} ,g^{(j)}_{\vk'}]
\!=\!
{\displaystyle \frac{\hbar }{\sqrt{N}}} \!
\left( \!
g^{(i)}_{\vk+\vk'}k_j
-
g^{(j)}_{\vk+\vk'}k'_i \!
\right) .
\EA
\label{CRgrho}  
\ea\\[-30pt]

Following Sunakawa
\cite{SYN2.62}, 
we define the modified momentum density operator
$\vv_{\vk}$ by\\[-24pt]
\ba
\BA{c}
\vv_{\vk}
\!\equiv
\vg_{\vk} 
\!-\!
{\displaystyle \frac{1}{ \sqrt {N}}}
\sum_{\vp \ne \vk} 
\rho_{\vp-\vk} \vv_{\vp}~ 
(\vk \!\ne\!  0) ,~~
\vv_0
\!=\!
0 .
\EA 
\label{modifiedv}  
\ea\\[-18pt]
Along the same way as the one in I and II and the Sunakawa's$\!$
\cite{SYN.62},
we prove two important CRs\\[-24pt] 
\ba
[\vv_{\!\vk}, \rho_{\!\vk'}]
\!=\! 
\hbar \vk' {\delta }_{\vk,\vk'} ,
\label{modifiedCRs}  
\ea
\vspace{-1.0cm}
\ba
\!\!\!\!\!\!
\BA{c}
[v^{(i)}_{\vk} , v^{(j)}_{\vk'}]
\!\approx\!
-
{\displaystyle \frac{\hbar}{N} } \!
\sum_{\vp\mbox{\scriptsize{all}}} \rho_{\vp-\vk-\vk'} \!
\left( \! p_i v^{(j)}_{\vp} \!-\! p_j v^{(i)}_{\vp} \! \right)
\!+\!
{\displaystyle \frac{1}{N} } \!
\sum'_{\vp,\vq} \rho_{\vp-\vk}\rho_{\vq-\vk'}
[v^{(i)}_{\vp} , v^{(j)}_{\vq}]  .
\EA 
\label{CRvv}  
\ea\\[-16pt]
The symbol $\!\sum'\!$ means that
the term
$\vp\!\!=\!\!\vk$ and $\vq\!\!=\!\!\vk'$$\!$
at the same time
should be omitted.
The CRs
(\ref{modifiedCRs})
and
(\ref{CRvv})
are quite the same as the Sunakawa's
\cite{SYN2.62}.
As pointed out by him,
Fourier transforms of the operators
$\rho_{\vk}$ and $\vv_{\vk}$
and the CRs among them
are identical with those found by Landau
\cite{Landau.71}
for the fluid dynamical density operator and the velocity operator.
Then, it turns out that the quantum mechanical operator
$\!\vv(\vx\!)$,
which satisfies the famous CR\\[-24pt]
\ba
\BA{c}
[v^{(i)}(\vx) , v^{(j)}(\vx')]
\!=\!
{\displaystyle \frac{i\hbar}{m}} 
\delta(\vx-\vx')
\rho(\vx)^{-1}(\mbox{rot}\vv(\vx))^{(k)}~
(i,~j,~k)~\mbox{cyclic},
\EA 
\label{CRvivj}  
\ea\\[-20pt]
corresponds to the fluid dynamical velocity.
We also have 
$
[\vv(\vx), \rho(\vx')]
\!=\!
-
{\displaystyle \frac{i\hbar}{m}}
\nabla_{\!x}
\delta(\vx-\vx').
$

\vspace{-0.4cm}


\def\thesection{\arabic{section}}
\setcounter{equation}{0}
\renewcommand{\theequation}{\arabic{section}.\arabic{equation}}     
\section{$\vv_{\vk}$- and $\rho_{\vk}$-dependence of the Hamiltonian}

~~~~We derive here a collective 
Hamiltonian in terms of the
$\vv_{\vk}$ and $\rho_{\vk}$.
Following Sunakawa, we expand
the kinetic operator $T$
in a power series of the velocity operator
$\vv_{\vk}$ 
as follows:\\[-22pt] 
\ba
\!\!
\BA{c}
T
\!\!=\!\! 
{T}_{0}(\rho)
\!\!+\!\! 
\sum_{\vp \ne 0} \!
{\T}_{1} (\rho ; \vp  ) \!\cdot\! \vv_{\vp}
\!\!+\!\!
\sum_{\vp \ne 0, \vq \ne 0} \!
{T}_{2} (\rho ; \vp,\vq )
\vv_{\vp} \!\cdot\! \vv_{\vq}
\!\!+\! 
\cdot \cdot \cdot  ,~
{T}_{2} (\rho ; \vp,\vq )
\!\!=\!\! 
{T}_{2} ( \rho ; \vq,\vp ) ,
\EA
\label{Texpansion}  
\ea\\[-22pt]
in which 
$\!{T_{n} (n \!\neq\! 0)}\!$ are unknown expansion coefficients. 
In order to determine their explicit expressions, we take
the CRs between $T$ and $\rho_{\vk}$
as follows:\\[-20pt]
\ba
\!\!
\BA{ll}
&
[T , \rho_{\vk}]
\!=\! 
\hbar
{\T}_{1} (\rho ; \vk  ) \!\cdot\! \vk
\!+\!
2 \hbar \!
\sum_{\vp \ne 0} \!
{T}_{2} (\rho ; \vp,\vk )
\vv_{\vp} \!\cdot\! \vk
\!+\! 
\cdots  ,~\\
\\[-8pt]
&
[[T, \rho_{\vk}] , \rho_{\vk'}]
\!=\!
2 \hbar^2 
{T}_{2} ( \rho ; \vk',\vk ) \vk' \!\cdot\! \vk
\!+\! 
\cdots  .
\EA \!\!
\label{Texpansioncommrho}  
\ea\\[-16pt]
From  
(\ref{FcomponentMomentumDensityOp}) and (\ref{modifiedv}),
we can calculate the CRs between $T$ and $\rho_{\vk}$
as follows:\\[-22pt]
\ba
\BA{ll}
&
[T , \rho_{\vk}]
\!=\!
{\displaystyle \frac{\hbar }{m}}
\vk
\!\cdot\!
\vg_{-\vk}
\!=\!
{\displaystyle \frac{\hbar }{m}}
\vk
\!\cdot\!
\vv_{-\vk}
\!+\!
{\displaystyle {\frac{\hbar}{m \! \sqrt {\!N}}}} \!\!
\sum_{\vp \ne -\vk} \!
\rho_{\vp+\vk}
\vk
\!\cdot\!
\vv_{\vp}  , \\
\\[-16pt]
&[[T, \rho_{\vk}] , \rho_{\vk'}]
\!=\!\!
-{\displaystyle \frac{ \hbar^{2} \! {\vk}^{2}}{m}} \! \delta_{\!\vk',-\vk } ,
{\displaystyle \frac{{\hbar}^{2}}{m \! \sqrt {\!N}}}
\vk
\!\cdot\!
\vk' \!
\rho_{\vk+\vk'} , \vk'  \!\!\ne\!\!  -\vk ,~~
[[[T , \rho_{\vk} ] , \rho_{\vk'}] , \rho_{\vk''}]
\!=\!
0 ,~
\!\cdots\! .
\EA \!\!
\label{CRTrhorho}
\ea\\[-12pt]
Comparing the above results with the CRs
(\ref{Texpansioncommrho}),
we can determine the coefficients 
${T_{n} (n \!\neq\! 0)}$. Then we can express the kinetic part ${T}$ in terms 
of the 
$\rho_{\vk}$ and $\vv_{\vk}$
as follows:\\[-22pt]
\ba
\BA{c}
T
\!=\!
{T}_{0}(\rho)
\!+\!
{\displaystyle \frac{1}{2m}} \!
\sum_{\vk \ne 0} \!
\vv_{\vk}
\!\cdot\!
\vv_{-\vk}
\!+\! 
{\displaystyle \frac{1}{2m \! \sqrt {\!N}}} \!
\sum_{\vp+\vq \ne 0} \!
\rho_{\vp+\vq}
\vv_{\vp}
\!\cdot\!
\vv_{\vq} ,~
(\vv_0 \!=\! 0) .
\EA
\label{exactTPi}  
\ea\\[-18pt]
Up to the present stage, all the expressions 
have been derived without any approximation. 
To determine 
$T_{0}(\rho)$,
we expand it in a power series of $\rho_{\vk}$
and take the CRs 
$[T_{0} (\rho), v^{(i)}_{\vk}]$
as follows:\\[-18pt] 
\ba
\!\!\!\!
\left.
\BA{ll}
&
{T}_{0} (\rho)
\!=\! 
{C}_{0}
+
\sum_{\vp \ne 0} \!
{C}_{1}(\vp) \rho_{\vp}
+
\sum_{\vp \ne 0, \vq \ne 0} \!
{C}_{2} (\vp,\vq)
\rho_{\vp} \rho_{\vq} 
+
\cdots  ,~
{C}_{2} (\vp,\vq ) \!=\!  {C}_{2} (\vq,\vp ) , \\
\\[-6pt]
&
[v^{(i)}_{\vk}  , {T}_{0}(\rho)]
\!=\! 
\hbar k_i
{C}_{1} (\vk  ) 
+
2 \hbar k_i \!
\sum_{\vp \ne 0} 
{C}_{2} (\vp ; \vk )
\rho_{\vp}
+ 
\cdots  ,~\\
\\[-6pt]
&
[v^{(j)}_{\vk'} , [v^{(i)}_{\vk}  , {T}_{0}(\rho)]]
\!=\!
2 \hbar^2 k_i k'_j
{C}_{2} ( \vk' ; \vk ) 
+
6 \hbar^2 k_i k'_j \!
\sum_{\vp \ne 0} 
{C}_{3} (\vp ; \vk' ; \vk ) 
\rho_{\vp}
+ 
\cdots  .
\EA \!\!
\right\}
\label{Texpansioncommvelo}  
\ea\\[-10pt]
From 
(\ref{modifiedv}),
we have a discrete integral equation\\[-22pt]
\ba
\!\!\!
\BA{c}
[v^{(i)}_{\vk}  , {T}_{0}(\rho)]
\!=\!
[g^{(i)}_{\vk} , {T}_{\!0}(\rho)]
\!-\!
{\displaystyle \frac{1}{\! \sqrt {N}}} \!
\sum_{\vp \ne \vk} \!
\rho_{\vp-\vk}
[v^{(i)}_{\vp}  , {T}_{0}(\rho)] ,
\left( \! \mbox{Denote}~
[g^{(i)}_{\vk} , {T}_{\!0}(\rho)]~
\mbox{as}~
f^{(i)}(\rho ; \vk) \!
\right) \!  .
\EA
\label{CRPiT0}  
\ea\\[-18pt]
With the aid of
(\ref{exactTPi})
and using two CRs of
(\ref{CRgrho}),
${f}^{(i)}(\rho ; \vk)$ is calculated approximately as\\[-20pt]
\ba
\!\!\!\!\!\!\!\!
\BA{ll}
&f^{(i)}(\rho ; \vk)
\!\approx\!
[ g^{(i)}_{\vk} , T]
\!-\! 
{\displaystyle \frac{1}{ 2m \! \sqrt {\!N}}} \!
\sum_{\vp+\vq \ne 0} 
[ g^{(i)}_{\vk} ,
\rho_{\vp+\vq}
] 
\vv_{\vp} \!\cdot\! \vv_{\vq} 
\!-\!
{\displaystyle \frac{1}{2m}} \!
\sum_{\vp \ne 0 } 
[ g^{(i)}_{\vk} ,    
\vv_{\vp} \!\cdot\! \vv_{-\vp}
] \\
\\[-12pt]
\!\!\!\!
&=\!
[g^{(i)}_{\vk} , T]
\!-\!
{\displaystyle \frac{\hbar }{m N}} \!
\sum_{\vp, \vq \mbox{\scriptsize all}} 
\rho_{\vp+\vq-\vk} \!
\left\{ \!
\left( \vk \!\cdot\! \vv_{\vp} \right) \!
 \vv_{\vq}^{(i)}
+
k_i 
\vv_{\vp} \!\cdot\! \vv_{\vq} \!
\right\} \\
\\[-12pt]
\!\!\!\!
&
-
{\displaystyle \frac{\hbar}{2 m \sqrt{N}}} \!
\sum_{\vp \!\ne\! 0 }  
\left\{  
(\vk \!\cdot\! \vv_{-\vp})
\vv_{\vk+\vp}^{(i)}
\!+\!
\vv_{\vk+\vp}^{(i)}
(\vk \!\cdot\! \vv_{-\vp})
\!-\!
p_i    
\vv_{-\vp} \!\cdot\! \vv_{\vk+\vp}
\!-\!
p_i    
\vv_{\vk+\vp} \!\cdot\! \vv_{-\vp}
\right\}  \\
\\[-12pt]
\!\!\!\!
&
=\!
[g^{(i)}_{\vk} , T]
\!-\!
\Lambda^{1(i)}_{\vk} 
\!-\!
\Lambda^{2(i)}_{\vk} 
\!-\!
\Lambda^{3(i)}_{\vk} ,~
\Lambda^{1(i)}_{\vk} 
\!\equiv\!
{\displaystyle \frac{\hbar }{m N}} \!
\sum_{\vp, \vq \mbox{\scriptsize all}} 
\rho_{\vp+\vq-\vk} \!
\left( \vk \!\cdot\! \vv_{\vp} \right) \!
 \vv_{\vq}^{(i)} ,
\EA
\label{fkrho}  
\ea\\[-12pt]
where\\[-24pt]
\ba
\!\!
\BA{l}
[g^{(i)}_{\vk} \! , T]
\!\!=\!\!
{\displaystyle \frac{ \hbar^{3}}{2m \! \sqrt {\!N}}} \!\!
\sum_{\!\vp~\!\!\mbox{\scriptsize all}} \!
p_i \!
\left\{ \!\!
\left( \! \vp \!+\! {\displaystyle \frac{\vk}{2}} \! \right)^{\!\!\!2}
\!\!-\!\!
\left( \! \vp \!-\! {\displaystyle \frac{\vk}{2}} \! \right)^{\!\!\!2} \!
\right\} \!
a^{~\!\!\dag}_{\!\vp-\frac{\vk}{2}} 
a_{\!\vp+\frac{\vk}{2}}
\!\!=\!\!
{\displaystyle \frac{ \hbar^{3}}{ m \! \sqrt {\!N}}} \!\!
\sum_{\!\vp~\!\!\mbox{\scriptsize all}} \!
p_i (\vp \!\cdot\! \vk) 
a^{~\!\dag}_{\!\vp-\frac{\vk}{2}} 
a_{\!\vp+\frac{\vk}{2}} ,
\EA
\label{commupiT}  
\ea\\[-30pt]
\ba
\!\!
\BA{l}
\Lambda^{2(i)}_{\vk}
\!\equiv\!
{\displaystyle \frac{\hbar }{m N}} \!
\sum_{\vp, \vq \mbox{\scriptsize all}} 
\rho_{\vp+\vq-\vk} \!
k_i 
\vv_{\!\vq} \!\cdot\! \vv_{\vp} ,
\EA
\label{fkrho1}  
\ea\\[-30pt] 
\ba
\!\!
\BA{l}
\Lambda^{3(i)}_{\vk}
\!\equiv\!
{\displaystyle \frac{\hbar}{2 m \sqrt{N}}} \!
\sum_{\vp \!\ne\! 0 } \! 
\left\{ \! 
(\vk \!\cdot\! \vv_{-\vp})
\vv_{\vk+\vp}^{(i)}
\!+\!
\vv_{\vk+\vp}^{(i)}
(\vk \!\cdot\! \vv_{-\vp})
\!-\!
p_i    
\vv_{-\vp} \!\cdot\! \vv_{\vk+\vp}
\!-\!
p_i    
\vv_{\vk+\vp} \!\cdot\! \vv_{-\vp}
\right\} .
\EA
\label{fkrho2}  
\ea\\[-26pt]

From now on, using
$\rho_{0} \!=\! \sqrt {N}$,
we make approximations for
$\vv_{\vk}, \rho_{\vk}, a_{0}$
and
$a^{\dag}_{0}$
as\\[-18pt]
\ba
\!\!\!\!\!\!
\BA{c}
\vv_{\vk}
\!\cong\!
{\displaystyle \frac{\hbar \vk}{2}} \!
\left( \!
\overline{\theta} a_{\vk}
\!-\!
a^{\dag}_{-\vk} \theta \!
\right) \! ,~
\rho_{\vk}
\!\cong\!
\overline{\theta} a_{-\vk}
\!+\!
a^{\dag}_{\vk} \theta ,~
a_{0} \!\cong\! \sqrt{N} \theta ,~
a^{\dag}_{0} \!\cong\! \sqrt{N} \overline{\theta} ,
\EA
\label{approxpirho}  
\ea\\[-14pt]
where
$\theta$ and $\overline{\theta}$ are the Grassmann numbers
\cite{Berezin.66}.
Using
$
\rho_{\vp+\vq-\vk}
\!\cong\!
\sqrt{N} \delta_{\vq,\vk-\vp}$,
$\Lambda^{1(i)}_{\vk}$ 
in 
(\ref{fkrho})
is computed as\\[-16pt]
\ba
\BA{ll}
\Lambda^{1(i)}_{\vk}
&\!\!\!=\!
{\displaystyle \frac{\hbar^3}{4 m \! \sqrt{N}}} \!
\sum_{\vp~\!\mbox{\scriptsize all}} 
\vk \!\cdot\! \vp  
(k_i \!-\! p_i) \!
\left( 
\overline{\theta} a_{\vk-\vp} 
\!-\!
a^{\dag}_{-(\vk-\vp)} \theta 
\right) 
\!\!
\left( 
\overline{\theta} a_{\vp} 
\!-\!
a^{\dag}_{-\vp} \theta 
\right) \\
\\[-10pt]
&\!\!\!=\!
{\displaystyle \frac{\hbar^3}{4 m \! \sqrt{N}}} \!
\sum_{\vp~\!\mbox{\scriptsize all}} 
\vp \!\cdot\! \vk 
(k_i \!-\! p_i) 
\rho_{-\vp}
\rho_{\!\vp \!-\! \vk\!} \\
\\[-10pt]
&~\!
+
\theta \overline{\theta}
{\displaystyle \frac{\hbar^3}{2 m \! \sqrt{N}}} \!
\sum_{\vp~\!\mbox{\scriptsize all}} \!
\left\{ \!
\left( \! \vp \!\cdot\! \vk \!-\! {\displaystyle \frac{\vk^2}{2}} \! \right) \!
\left( \! p_i \!+\!{\displaystyle \frac{ k_i}{2}} \! \right)  
\!+\!
\left( \! \vp \!\cdot\! \vk \!+\! {\displaystyle \frac{\vk^2}{2}} \! \right) \!
\left( \! p_i \!-\!{\displaystyle \frac{ k_i}{2}} \! \right) \!
\right\} \!
a^{\dag}_{\vp \!-\! \frac{\vk}{2}} 
a_{\vp \!+\! \frac{\vk}{2}} \\
\\[-10pt] 
&\!\!\!=\!
{\displaystyle \frac{\hbar^3}{4 m \! \sqrt{N}}} \!
\sum_{\vp \mbox{\scriptsize all}} 
\vp \!\cdot\! \vk 
(k_i \!\!-\!\! p_i) 
\rho_{-\vp}
\rho_{\!\vp \!-\! \vk\!}
+
\theta \overline{\theta} 
{\displaystyle \frac{\hbar^3}{m }} \!
\left( \!
{\displaystyle \frac{\vk}{2}} \!\cdot\! \vk
{\displaystyle \frac{k_i}{2}}
\!-\! 
{\displaystyle \frac{\vk^2}{4}} k_i \!
\right) \!
\rho_{\vk} , \\
\EA
\label{approxrhokvv}  
\ea
where the last term in the last line is obtained by extracting the terms with
$\!\vp \!\!=\!\! {\displaystyle \frac{\vk}{2}}\!$
or
$\!\vp \!\!=\!\! -{\displaystyle \frac{\vk}{2}}\!$
in the $\theta \overline{\theta}$ term of the third line.
As is clear from its structure, it evidently vanishes.
The remaining terms in the $\theta \overline{\theta}$ term
are small and can be neglected.
The terms
$\!\Lambda^{2(i)}_{\vk}\!$
and
$\!\Lambda^{3(i)}_{\vk}\!\!$
in 
(\ref{fkrho1}) and (\ref{fkrho2}),
which have never been seen in the previous paper
\cite{NishProvi2.16},
are also calculated similarly as the above,
respectively
and are given as\\[-18pt]
\ba
\!\!
\BA{ll}
\Lambda^{2(i)}_{\vk}
&\!\!\!\!=\!
{\displaystyle \frac{\hbar^3 k_i}{4 m \! \sqrt{N}}} \!
\sum_{\vp~\!\mbox{\scriptsize all}} 
(\vk \!-\! \vp) \!\cdot\! \vp
\left( 
\overline{\theta} a_{\vk-\vp} 
\!-\!
a^{\dag}_{-(\vk-\vp)} \theta 
\right) 
\!\!
\left( 
\overline{\theta} a_{\vp} 
\!-\!
a^{\dag}_{-\vp} \theta 
\right) \\
\\[-10pt]
&\!\!\!\!=\!
-{\displaystyle \frac{\hbar^3 k_i}{4 m \! \sqrt{N}}} \!\!
\sum_{\vp \mbox{\scriptsize all}} 
\vp \!\cdot\! (\vp \!-\! \vk) 
\rho_{-\vp}
\rho_{\!\vp \!-\! \vk\!}
-
\theta \overline{\theta} 
{\displaystyle \frac{3 \hbar^3 \vk^2  k_i }{16 m }} 
\rho_{-\vk} ,
\EA
\label{approxrhokvv2}  
\ea\\[-30pt]
\ba
\!\!
\BA{ll}
\Lambda^{3(i)}_{\vk}
&\!\!\!\!=\!
-{\displaystyle \frac{\hbar^3 }{4 m \! \sqrt{N}}} \!
\sum_{\vp \!\ne\! 0} \!
\left\{ 
( k_i \!+\! p_i)  
\vk \!\cdot\! \vp 
\!-\!
p_i  
\left( \vk \!+\! \vp \right) \!\cdot\! \vp 
\right\} \!\!
\left( 
\overline{\theta} a_{\vk+\vp} 
\!-\!
a^{\dag}_{-(\vk+\vp)} \theta 
\right) 
\!\!
\left( 
\overline{\theta} a_{\vp} 
\!-\!
a^{\dag}_{-\vp} \theta 
\right) \\
\\[-10pt]
&\!\!\!\!=\!
-{\displaystyle \frac{\hbar^3 }{4 m \! \sqrt{N}}} \!\!
\sum_{\vp \!\ne\! 0} \!
\left\{ 
( k_i \!+\! p_i) 
\vk \!\cdot\! \vp 
\!-\!
p_i  
\left( \vk \!+\! \vp \right) \!\cdot\! \vp
\right\} \!
\rho_{\vp}
\rho_{-(\vp \!+\! \vk)} .
\EA
\label{approxrhokvv3}  
\ea\\[-10pt]
Here
in
(\ref{approxrhokvv2})
we use the relation
$\theta \overline{\theta} \!=\! 1$.
 
Substituting
$
[g^{(i)}_{\vk}, T]
\!\!\cong\!\!
{\displaystyle \frac{{\hbar }^{3} \! {\vk}^{2} \! k_i }{4m}} \! 
\rho_{-\vk}
$
and
$\Lambda^{(i)}_{\vk}$'s,
we get an approximate formula for
${{f}^{(i)}(\rho ; \vk)}$
as\\[-10pt]
\ba
\!\!\!\!
\BA{c}
{f}^{(i)} \! (\!\rho ; \vk\!)
\!\!=\!\!
{\displaystyle {\frac{7{\hbar }^{3}{\vk}^{2} k_i }{16m}}} \!\!
\rho_{\!-\vk}
\!-\!
{\displaystyle \frac{\hbar^3}{4 m \! \sqrt{\!N}}} \!\!
\sum_{\!\vp \!\ne\! \vk} \!\!
\left\{ 
\vp \!\cdot\! \vk 
(\!k_i \!\!-\!\! p_i\!)
\!\!-\!\!
k_i \vp \!\cdot\! (\!\vp \!\!-\!\! \vk\!)
\!\!-\!\! 
(\! k_i \!\!+\!\! p_i\!) 
\vk \!\cdot\! \vp
\!\!+\!\!
p_i \! 
\left(\! \vk \!\!+\!\! \vp \!\right) \!\cdot\! \vp 
\right\} \!\!
\rho_{\!-\vp} 
\rho_{\!\vp \!-\! \vk} \\
\\[-6pt]
 \!=\!
{\displaystyle \frac{7}{4}}
{\displaystyle {\frac{{\hbar }^{3}{\vk}^{2} k_i }{4m}}} 
\rho_{-\vk}
\!-\!
{\displaystyle \frac{\hbar^3}{4 m  \sqrt{N}}} 
\sum_{\vp \!\ne\! \vk} 
\left\{ 
(\vp \!\cdot\! \vk)
(k_i \!-\! p_i)
\!-\!
\vp^2
(k_i \!-\! p_i)
\right\} 
\rho_{-\vp} 
\rho_{\vp \!-\! \vk} .
\EA
\label{approxfkrho}  
\ea\\[-8pt]
In the last line,
the numerical factor
${\displaystyle \frac{7}{4}}$
and the term
$\!-\! \vp^2 (k_i \!-\! p_i)$,
which have also never been seen in the previous paper
\cite{NishProvi2.16},
arise due to the consideration of
the terms
$\!\Lambda^{2(i)}_{\vk}\!$
and
$\!\Lambda^{3(i)}_{\vk}\!\!$.
Further substituting 
(\ref{approxfkrho}) into 
(\ref{CRPiT0})
and picking up the next leading term,
we have\\[-16pt]
\ba
\!\!\!\!
\BA{l} 
[v^{(i)}_{\vk} , {T}_{0} (\rho)]
\!\approx\! 
{\displaystyle {\frac{7{\hbar }^{3} {\vk}^{2} k_i }{16m}}} \!
\rho_{-\vk}
\!-\!
{\displaystyle \frac{\hbar^3 }{4 m \! \sqrt{N}}} \!
\sum_{\vp \ne \vk} \!
\left\{ \!
(\vp \!\cdot\! \vk \!-\! \vp^2)
(k_i \!-\! p_i)
\!+\!
{\displaystyle \frac{7}{4}}
\vp^2 p_i \!
\right\} \!
\rho_{-\vp}
\rho_{\vp \!-\! \vk} .
\EA
\label{CRPiT02}  
\ea\\[-10pt]
To keep a favored form
$\sum_{\vp \ne \vk}  
(\vp \!\cdot\! \vk) (k_i \!-\! p_i)  
\rho_{-\vp} \rho_{\vp \!-\! \vk}
$
occurred in
\cite{NishProvi2.16},
in the curly brackets,
we had better to treat
$k_i \!-\! p_i$ and ${\displaystyle \frac{7}{4}}p_i$
equally
and to lead
$\vp \!\cdot\! (\vp \!+\! \vk)$.
So, we take
$p_i \!=\! {\displaystyle \frac{8 }{15}}k_i$
and then
$k_i \!-\! p_i \!=\! {\displaystyle \frac{7 }{15}}k_i$.
Thus, we have a final expression for
$[v^{(i)}_{\vk} , {T}_{0} (\rho)]$
up to the order of ${\displaystyle \frac{1}{\sqrt{N}}}$
as\\[-14pt]
\ba
\!\!\!\!
\BA{l}
[v^{(i)}_{\vk} , {T}_{0} (\rho)] 
\!=\!  
{\displaystyle \frac{7}{4}}
{\displaystyle {\frac{{\hbar }^{3}{\vk}^{2} k_i }{4m}}} \!
\rho_{-\vk}
\!-\!
{\displaystyle \frac{14}{15}}
{\displaystyle \frac{\hbar^3 k_i }{8 m \sqrt{N}}} \!
\sum_{\vp \ne \vk} 
\vp \!\cdot\! (\vp \!+\! \vk)
\rho_{-\vp}
\rho_{\vp \!-\! \vk} .
\EA
\label{CRPiT03}  
\ea\\[-12pt]
From
(\ref{CRPiT02})
and 
(\ref{modifiedCRs}),
we get the following CRs
between
$v^{(i)}_{\vk}$ and ${T}_{\!0} (\rho)$:\\[-16pt] 
\ba
\left.
\BA{ll}
&
[v^{(j)}_{\vk'} \! , [v^{(i)}_{\vk} \! , {T}_{\!0} (\rho)]]
\!=\! 
-
{\displaystyle \frac{7}{4}} 
{\displaystyle {\frac{{\hbar }^{4} k_i k_j }{4m}}}
{\vk}^{2}{\delta }_{\vk', -\vk} 
\!-\!
{\displaystyle \frac{14}{15}} 
{\displaystyle {\frac{{\hbar }^{4} k_i k'_j}{4 m \sqrt {N}}}} \!
\left( 
\vk^2
\!+\!
\vk \!\cdot\! \vk'
\!+\!
\vk'^2 
\right) \!
\rho_{-\vk-\vk'} , \\
\\[-4pt]
&
[v^{(k)}_{\vk''} \! , [v^{(j)}_{\vk'} \! , [v^{(i)}_{\vk} \! , {T}_{\!0}(\rho)]]]
\!=\! 
{\displaystyle \frac{14}{15}} 
{\displaystyle
{\frac{{\hbar }^{5} k_i k'_j (k_k \!+\! k'_k)}{4 m \sqrt {N}}}
} \!
\left( 
\vk^2
\!+\!
\vk \!\cdot\! \vk'
\!+\!
\vk'^2 
\right)  \! , \\
\\[-4pt]
&
[v^{(l)}_{\vk'''} , [v^{(k)}_{\vk''}, [v^{(j)}_{\vk'} , [v^{(i)}_{\vk} , 
{T}_{\!0} (\rho )]]]]
\!=\!  
0  .
\EA \!\!\!\!
\right\}
\label{CRPiPiPiPiT0}  
\ea\\[-10pt]
By the procedure similar to the previous one,
we can determine the coefficients
${C_{2}}$
and
${C_{3}}$
in
(\ref{Texpansioncommvelo})
as
$
{C}_{2} ( \vk' ; \vk ) 
\!=\!
{\displaystyle \frac{7}{4}}
{\displaystyle {\frac{{ \hbar }^{2}{\vk}^{2}}{8m}}}
\delta_{\vk', -\vk}
$
and
$
{C}_{3} (\vk'' ; \vk' ; \vk )
\!=\!
-{\displaystyle \frac{7}{15}}
{\displaystyle {\frac{{ \hbar }^{2}}{12m \sqrt{N}}}} \!
\left( \!
{\vk}^{2}
\!+\!
\vk \!\cdot\! \vk'
\!+\!
{\vk'}^{2} \!
\right) \!
\delta_{\vk'',-\vk -\vk'}
$
which are the same forms as the Sunakawa's
${C_{2}}$
and
${C_{3}}$
\cite{SYN2.62}
but having different numerical factors. 
This is also due to the consideration of
the terms
$\!\Lambda^{2(i)}_{\vk}\!$
and
$\!\Lambda^{3(i)}_{\vk}\!\!$.
Then get an approximate form of ${T_{0}(\rho)}$
in terms of variables $\rho_{\vk}$ as\\[-20pt]
\ba
\!\!\!\!
\BA{c}
{T}_{\!0} (\rho)
\!\!=\!\!
{C}_{\!0} 
\!+\!
{\displaystyle \frac{7}{4}}
{\displaystyle {\frac{{ \hbar }^{2}}{8m}}} \!\!
\sum_{\!\vk \ne 0} \!
{\vk}^{2} \!
\rho_{\!\vk} \rho_{\!-\vk}
\!-\!
{\displaystyle \frac{14}{15}}
{\displaystyle {\frac{{\hbar }^{2}}{24 m \! \sqrt {\!N}}}} \!\!
\sum_{\!\vp \ne 0, \vq \ne 0, \vp+\vq \ne 0} 
(\vp^2 \!+\! \vp \!\cdot\! \vq \!+\! \vq^2)
\rho_{\!\vp} \rho_{\!\vq} \rho_{\!-\vp-\vq} .
\EA
\label{T0rho2}  
\ea\\[-16pt]
Substituting
(\ref{T0rho2})  
into
(\ref{exactTPi})
and
using the underlying identity,\\[-20pt]
\ba
\BA{c}
\sum_{\vp \ne 0, \vq \ne 0, \vp+\vq \ne 0} 
(\vp^2 \!+\! \vp \!\cdot\! \vq \!+\! \vq^2)
\rho_{\vp} \rho_{\vq} \rho_{-\vp-\vq} 
\!=\!
-3 \!
\sum_{\vp \ne 0, \vq \ne 0, \vp+\vq \ne 0} 
\vp \!\cdot\! \vq 
\rho_{\vp} \rho_{\vq} \rho_{-\vp-\vq}   ,
\EA
\label{identity}  
\ea\\[-18pt]
the constant term $C_{0}$ is computed
as\\[-18pt]
\ba
\BA{lll}
{C}_{0} 
=
T
\!\!\!\!&-&\!\!\!\! 
{\displaystyle \frac{7}{4}}
{\displaystyle {\frac{{\hbar }^{2}}{8m}}} \!
\sum_{\vk \ne 0} 
{\vk}^{2} \!
\rho_{\vk} \rho_{-\vk} 
\!-\! 
{\displaystyle {\frac{1}{2m}}} \!
\sum_{\vk} \!
\vv_{\vk} \!\cdot\! \vv_{-\vk} 
\!-\!
{\displaystyle {\frac{1}{2 m \! \sqrt {N}}}} \!
\sum_{\vp+\vq \ne 0} \!
\rho_{\vp+\vq} \vv_{\vp} \!\cdot\! \vv_{\vq} \\
\\[-12pt]
\!\!\!\!&-&\!\!\!\!
{\displaystyle \frac{14}{15}}
{\displaystyle {\frac{{\hbar }^{2}}{8 m \! \sqrt {N}}}}
\sum_{\vp \ne 0, \vq \ne 0, \vp+\vq \ne 0} 
\vp  \!\cdot\! \vq
\rho_{\vp} \rho_{\vq} \rho_{-\vp-\vq} .
\EA
\label{C0phiexpansion}
\ea\\[-10pt]
Using
(\ref{approxpirho}),
we can calculate the third term in
(\ref{C0phiexpansion}) 
and then reach to a result such as \\[-18pt]
\ba
\BA{l}
- 
{\displaystyle {\frac{1}{2m}}} \!
\sum_{\vk} 
\vv_{\vk} \!\cdot\! \vv_{-\vk} 
\!=\!
{\displaystyle \frac{{\hbar }^{2}}{8 m}} \!
\sum_{\vk} 
\vk^{2} \!
\left( 
\overline{\theta} a_{\vk} 
\!-\!
a^{\dag}_{-\vk} \theta 
\right) \!\!
\left( 
\overline{\theta} a_{-\vk} 
\!-\!
a^{\dag}_{\vk} \theta 
\right) \\
\\[-6pt] 
\!=\! 
{\displaystyle \frac{{\hbar }^{2}}{8 m}} \!
\sum_{\vk}  
\vk^{2}
\rho_{\vk} \rho_{-\vk}
\!+\!
\theta \overline{\theta}
{\displaystyle \frac{{\hbar }^{2}}{2 m}} \!
\sum_{\vk} 
\vk^{2}
\!-\! 
\theta \overline{\theta} 
\sum_{\vk} 
{\displaystyle \frac{{\hbar }^{2} \vk^{2}}{2 m}} \! 
a^{\dag}_{\vk} 
a_{\vk} .
\EA
\label{approxPiPi2}  
\ea\\[-8pt] 
As for the forth and last terms in
(\ref{C0phiexpansion}),
they become vanishing
due to 
$\theta \theta \!=\! 0$
and
$\overline{\theta} ~\! \overline{\theta} \!=\! 0$.
Substituting these into
(\ref{C0phiexpansion})
and replacing
$\rho_{\vk} \rho_{-\vk}$
by
$<\!\!\rho_{\vk} \rho_{-\vk}\!\!>_{\!Ave}$, 
we have \\[-14pt]
\ba
\!\!\!\!\!\!
\BA{l}
C_{0}
\!=\!
\left( 1 \!-\! \theta\overline{\theta} \right) \!
T
\!+\!
\theta\overline{\theta} 
{\displaystyle {\frac{{\hbar }^{2}}{2m}}} \!\!
\sum_{\vk} \! {\vk}^{2}
\!\!-\!
{\displaystyle {\frac{3{\hbar }^{2}}{32m}}} \!\!
\sum_{\vk } \!
{\vk}^{2} \!
\rho_{\vk} \rho_{-\vk}
\!\!\cong\!\!
\sum_{\vk} \!
{\displaystyle {\frac{{\hbar }^{2}{\vk}^{2}}{2 m}}} 
 \!-\!
{\displaystyle {\frac{3{\hbar }^{2}}{32m}}} \!\!
 \sum_{\vk } \!
{\vk}^{2} \!\!
<\!\!\rho_{\vk} \rho_{-\vk}\!\!>_{\!Ave} ,
\EA
\label{resultC0}
\ea\\[-10pt]
where the quantity
$<\!\!\rho_{\vk} \rho_{-\vk}\!\!>_{\!Ave}$
stands for the average value of
$\rho_{\vk} \rho_{-\vk}$  
and we have used the relation
$\theta \overline{\theta} \!=\! 1$.
It is surprising to see that the $C_{0}$
coincides with the constant term in the
ground-state energy
given by Tomonaga
\cite{Tomo.50}
except the last term. 
Using
(\ref{exactTPi}), (\ref{T0rho2}) and (\ref{resultC0}),
we can express the Hamiltonian
$H(=H^{\!\mbox{I}} \!+\! H^{\!\mbox{I\!I}})$
in terms of
$\rho_{\vk}$ and $\vv_{\vk}$
as follows:\\[-14pt]
\ba
\!\!\!\!\!\!
\left.
\BA{lll}
H^{\!\mbox{I}} 
&&\!\!\!\!\!\!\!\!\!
\!=\!
{\displaystyle \frac{N \!\! \left(\!N \!\!-\!\! 1 \!\right)}{2\Omega}} \!
\nu(0) 
\!\!+\!\!
\sum_{\! \vk \!\ne\! 0} \!\!
\left[ \!
- 
{\displaystyle \frac{\sqrt{7}}{4}\frac{{\hbar }^{2} \! {\vk}^{2}}{2m}}
\!\!-\!\!
{\displaystyle \frac{\sqrt{7}}{7}\frac{N}{\Omega}} \!
\nu(\vk)
\!\!+\!\!
{\displaystyle \frac{1}{2m}} 
{v}_{\vk} \!\cdot\! {v}_{-\vk}
\!\!+\!\!
\left( \!
{\displaystyle 
\frac{7{\hbar }^{2} \! {\vk}^{2}}{32 m}}
\!\!+\!\!
{\displaystyle \frac{N}{2\Omega}} \!
\nu(\vk) \!\!
\right) \!\!
\rho_{\!\vk} \rho_{\!-\vk} \!
\right]   \\
\\[-12pt]
~~~~ 
&&\!\!\!\!\!\!\!\!~~~
-
{\displaystyle {\frac{3{\hbar }^{2}}{32m}}} \!\!
 \sum_{\vk } \!
{\vk}^{2} \!\!
<\!\!\rho_{\vk} \rho_{-\vk}\!\!>_{\!Ave}
\!+\!
\left( \! 1 \!+\! {\displaystyle \frac{ \sqrt{7}}{4}} \! \right) \!
\sum_{\vk} \!
{\displaystyle \frac{{\hbar }^{2}{\vk}^{2}}{2m}}
\!-\!
\left( \! 1 \!-\! {\displaystyle \frac{2 \sqrt{7}}{7}} \! \right) \!
{\displaystyle \frac{N}{2\Omega}} \!
\sum_{\vk \ne 0} 
\nu (\vk) , \\
\\[-6pt]
H^{\!\mbox{I\!I}} 
&&\!\!\!\!\!\!\!\!
\!=\!
{\displaystyle \frac{1}{2 m \! \sqrt {\!N}}} \!\!
\sum_{\!\vp \ne 0, \vq \ne 0, \vp\!+\!\vq \ne 0} \!
\rho_{\!\vp+\vq}{\vv}_{\vp} \!\cdot\! {\vv}_{\vq}
\!+\! 
{\displaystyle \frac{14}{15}}
{\displaystyle {\frac{{\hbar }^{2}}{8 m \! \sqrt {\!N}}}} \!\!
\sum_{\!\vp \ne 0, \vq \ne 0, \vp\!+\!\vq \ne 0} \!
\vp \!\cdot\! \vq
\rho_{\vp} \rho_{\!-\vp-\vq} \rho_{\vq} ,
\EA \!\!\!\!
\right\} 
\label{exactH}
\ea\\[-8pt]
which is similar to the Sunakawa's Hamiltonian
\cite{SYN.62}
except the term
$
<\!\!\rho_{\vk} \rho_{-\vk}\!\!>_{\!Ave} 
$
in $H^{\!\mbox{I}}$.

Now, let us introduce
the Boson annihilation and creation operators
defined as\\[-14pt]
\ba
\BA{c}
\alpha_{\vk}
\!\equiv\!
\sqrt{\!\displaystyle{\frac{ m E_{\vk}}{2{\hbar }^{2}{\vk}^{2}}}} 
\rho_{-\vk}
\!+\!
\displaystyle{\frac{1}{\sqrt{\!2 m {\vk}^{2} \! E_{\vk}}}}
\vk \!\cdot\! \vv_{\vk},~
\alpha^{\dag}_{\vk}
\!\equiv\!
\sqrt{\!\displaystyle{\frac{ m E_{\vk}}{2{\hbar }^{2} {\vk}^{2}}}} 
\rho_{\vk}
\!+\!
\displaystyle{\frac{1}{\sqrt{\!2 m {\vk}^{2} \! E_{\vk}}}}
\vk \!\cdot\! \vv_{-\vk},~\!
(\vk \!\ne\! 0) .
\EA
\label{Boson_ops}
\ea\\[-12pt]
Using
(\ref{Boson_ops}),
$
\lambda_{\!\vk} 
\!\!\equiv\!\!
{\displaystyle \frac{{\hbar }^{2}{\vk}^{2}}{2 m E_{\vk}}}
$
and
(\ref{approxpirho}),
the collective variables
$\rho_{\!\vk}$ and $\vv_{\!\vk} (\vk \!\ne\! 0) $
are expressed as\\[-16pt]
\ba
\BA{l}
\rho_{\vk}
\!=\!
\sqrt{\lambda_\vk} \!
\left( \!
\alpha_{-\vk}
\!+\!
\alpha^\dag_{\vk} \!
\right)
\!\cong\! 
\overline{\theta} a_{-\vk}
\!+\!
a^{\dag}_{\vk} \theta ,~
\vv_{\vk}
\!=\!
{\displaystyle \frac{\hbar \vk}{2 \! \sqrt{\lambda_\vk}}} \!
\left( \!
\alpha_{\vk}
\!-\!
\alpha^\dag_{-\vk} \!
\right)
\!\cong\!
{\displaystyle \frac{\hbar \vk}{2}} \!
\left( \!
\overline{\theta} a_{\vk}
\!-\!
a^{\dag}_{-\vk} \theta \! 
\right) . 
\EA
\label{rhoPi}
\ea\\[-16pt]
The
$H^{\!\mbox{I}}$
in
(\ref{exactH})
is considered as the lowest order Hamiltonian
within the scope of the present approximation.
With the use of an inverse transformation of
(\ref{Boson_ops}),
the lowest order Hamiltonian
$H^{\!\mbox{I}}$
denoted as
$H_{0}$
is diagonalized as
\\[-18pt]
\ba
\!\!\!\!
\left.
\BA{l}
~\!H_0
\!=\!
E^G_0
\!+\!
\sum_{\vk \ne 0} \! E_{\vk}
\alpha^{\dag}_{\vk} \alpha_{\vk} , \\
\\[-12pt]
E^G_0
\!\equiv\!
{\displaystyle \frac{N  \left(N \!-\! 1 \right)}{2\Omega}}  
\nu (0)
\!+\!
{\displaystyle \frac{1}{2}} \!
\sum_{\vk \ne 0} \!
\left( \!
E_{\vk}
\!-\! 
{\displaystyle \frac{\sqrt{7}{\hbar }^{2} {\vk}^{2}}{4m}}
\!-\!
{\displaystyle \frac{2\sqrt{7} N}{7 \Omega}} 
\nu (\vk) 
\!-\!
{\displaystyle {\frac{3{\hbar }^{2}{\vk}^{2}}{16m}}} \!
<\!\!\rho_{\vk} \rho_{-\vk}\!\!>_{\!Ave} \!
\right) \! ,\\
\\[-14pt]
E_{\vk}
\!\equiv\!
\sqrt{ \!
{\displaystyle \frac{7}{4}} 
\varepsilon_{\vk}^{2}
\!+\!
{\displaystyle \frac{{\hbar }^{2} {\vk}^{2}}{m} 
\frac{N}{\Omega}} 
\nu(\vk)} ,~
\varepsilon_{\vk}
\!\equiv\!
{\displaystyle \frac{{\hbar }^{2} {\vk}^{2}}{2m}} ,
\EA \!\!\!
\right\}
\label{diagoH0}
\ea\\[-14pt]
where
$E_{\vk}$ is the quasi-particle energy
but with the modified single-particle energy
${\displaystyle \frac{\sqrt{7}}{2}}\varepsilon_{\vk}$
in the lowest approximation
and
we have used
$\!
[\vv_{\!\vk}, \rho_{\!\vk}]
\!\!=\!\! 
\hbar \vk 
\!$.
In this sense,
the zero point energy of the collective mode
is included in the above diagonalization.
The quantity $E^G_0$ in 
(\ref{diagoH0})
corresponds to the lowest order ground-state energy
\cite{BZ.55,Bogoliubov.47}
but with the  average value
$<\!\!\rho_{\vk} \rho_{-\vk}\!\!>_{\!Ave}$.
The part of the Hamiltonian, $H^{\!\mbox{I\!I}}$,
is also represented in terms of the Boson operators
$\alpha_\vk$ and $\alpha^\dag _\vk$
though we omit here its explicit expression.
Further
we have a Bogoliubov transformation
for Boson-like operators
$\overline{\theta} a_{\vk}$
and
$a^{\dag}_{\vk} \theta$
as \\[-20pt]
\ba
\!\!\!\!\!\!
\BA{l}
\alpha_{\vk}
\!=\!
{\displaystyle
\frac{ \left( E_{\vk} \!+\! \varepsilon_{\vk} \right) 
\overline{\theta} a_{\vk}
\!+\!
\left( E_{\vk} \!-\! \varepsilon_{\vk} \right) 
a^{\dag}_{-\vk} \theta  }
{2 \sqrt{ \varepsilon_{\vk} E_{\vk}}}
} , ~
\alpha^{\dag}_{-\vk}
\!=\!
{\displaystyle
\frac{ \left( E_{\vk} \!-\! \varepsilon_{\vk} \right) 
\overline{\theta}a_{\vk}
\!+\!
\left( E_{\vk} \!+\! \varepsilon_{\vk} \right)
a^{\dag}_{-\vk} \theta  }
{2 \sqrt{ \varepsilon_{\vk} E_{\vk}}}
} ,
\EA
\label{Bogolon_ops}
\ea\\[-14pt]
which already appeared in the preceding papers
\cite{NishProvi.15,NishProvi2.16}
and is of
the same form as the Bogoliubov transformation
for the usual Bosons
\cite{Bogoliubov.47}
except the introduction of  the Grassmann numbers
$\theta$ and $\overline{\theta}$.
The above kind of the diagonalization
(\ref{diagoH0})
has also been
given by Sunakawa in the Boson system
\cite{SYN3.62}.



\def\thesection{\arabic{section}}
\setcounter{equation}{0}
\renewcommand{\theequation}{\arabic{section}.\arabic{equation}}
\section{Discussions and further outlook}

~~~
In this paper,
we have proposed a velocity operator approach
to a 3D Fermion system.
After introducing collective variables,
the velocity operator approach
to the 3D Fermion system could be provided.
Particularly,
an interesting problem of describing
elementary excitations occurring in a strongly correlated system
may be possibly treated as an elementary exercise.
For such problems, for example,
see textbooks
\cite{RS.80,NO.88}.
By applying the velocity operator approach to such a problem,
an excellent description of the elementary excitations
in the 3D Fermion system will be expected
to reproduce various possible behaviors including excited energies.
Because the present theory is constructed to take into account
important many-body correlations which
have not been investigated sufficiently
for a long time 
in the historical ways for such a problem.
Here the velocity operator
$\vv^{(i)}_{\vk}$ is defined through the discrete integral equation
which is necessarily accompanied by the inhomogeneous term
${f}^{(i)}(\rho ; \vk)$.
It is a very important task how to evaluate this term
which brings an essential effect to
the construction of the theory for the velocity operator approach.
More precisely,
in the last line of
(\ref{fkrho}),
we have taken into account the contributions from
all of the second term $\Lambda^{1(i)}_{\vk}$,
the third one $\Lambda^{2(i)}_{\vk}$ and 
the forth one $\Lambda^{3(i)}_{\vk}$.
Such an evaluation is quite different from the manner
adopted in the preceding papers
\cite{NishProvi.15,NishProvi2.16}
in which only the term $\Lambda^{1(i)}_{\vk}$
had been taken into account.
As a result,
the present evaluation leads to another approximate formula for
${f}^{(i)}(\rho ; \vk)$
and another final form of
$[v^{(i)}_{\vk} , {T}_{0} (\rho)]$
which are of course different from those given
in those papers.
Under the use of the CRs
(\ref{CRPiPiPiPiT0})
between
$v^{(i)}_{\vk}$ and ${T}_{0} (\rho)$,  
thus,  the correct expression for ${T}_{0} (\rho)$
could be determined in terms of $\rho_{\vk}$
and the constant $C_{0}$
appearing in the expansion of
 a modified ${T}_{0} (\rho)$
turns out to involve the term
$<\!\!\rho_{\vk} \rho_{-\vk}\!\!>_{\!Ave}$
which is not determined yet.
However, now, making use of the inverse transformation of
(\ref{Boson_ops}), 
we determine a notable term as
$
<\!\!\rho_{\vk} \rho_{-\vk}\!\!>_{\!Ave}
=\!
{\displaystyle  \frac{{\hbar }^{2} {\vk}^{2}}{2mE_{\vk}}} \!\!
<\!\!
(\alpha_{-\vk} \!+\! \alpha^{\dag}_{\vk})
(\alpha_{\vk} \!+\! \alpha^{\dag}_{-\vk})
\!\!>_{\!Ave}
=\!
{\displaystyle \frac{{\hbar }^{2} {\vk}^{2}}{2mE_{\vk}}}
\!=\!
{\displaystyle \frac{\varepsilon_{\vk}}{E_{\vk}}}
$.
Using the last two energies
$E_{\vk}$ and $\varepsilon_{\vk}$ in
(\ref{diagoH0}),
here we approximate
$E_{\vk}$
and the term
$
-
{\displaystyle \frac{3{\hbar }^{2}{\vk}^{2}}{16m}} \!\!
<\!\!\rho_{\vk} \rho_{-\vk}\!\!>_{\!Ave}
$ in $E^G_0$ of
(\ref{diagoH0})
as
${\displaystyle \frac{\sqrt{7}}{2}}\varepsilon_{\vk}$
and
$
-
{\displaystyle \frac{3\sqrt{7}}{28}} \!\!
\varepsilon_{\vk}
$,
respectively.
Then the lowest order ground-state energy
$E^G_0$
is rewritten as
$
E^G_0
\!\!=\!\!
{\displaystyle \frac{N \! \left(\! N \!\!-\!\! 1 \!\right)}{2\Omega}}  
\nu (0)
\!+\!
{\displaystyle \frac{1}{2}} \!\!
\sum_{\vk \!\ne\! 0} \!\!
\left( \!\!
E_{\vk}
\!-\! 
{\displaystyle \frac{17\sqrt{7}}{28}} 
\varepsilon_{\vk}
\!-\!
{\displaystyle \frac{2\sqrt{7}}{7 }}
{\displaystyle \frac{N}{\Omega}} 
\nu (\vk) \!\!
\right) \!
$.
Further
we derive the Bogoliubov transformation
for Boson-like operators
$\overline{\theta} a_{\vk}$
and
$a^{\dag}_{\vk} \theta$
which already appeared in the preceding papers
\cite{NishProvi.15,NishProvi2.16}
and is the same form as the famous Bogoliubov transformation
for the usual Bosons
\cite{SYN2.62}
but which brings the modified single-particle energy
${\displaystyle \frac{\sqrt{7}}{2}}\varepsilon_{\vk}$
in the quasi-particle excitation energy $E_{\vk}$
in the lowest order approximation.
Thus, we could provide the modified and then the correct
theory for the velocity operator approach.
Based on such approach,
it is expected that
a new field of exploration of elementary excitation
in a 3D Fermi system
may open up
with aid of the velocity operator approach
whose new development may appear elsewhere.
The connection of the present theory with
the fluid dynamics was been mentioned briefly
by Sunakawa.
Restrict the Hilbert space to a subspace
in which the vortex operator satisfies
$\mbox{rot}\vv(\!\vx\!)| \!\! >=\!0$,
$(\vk \!\times\! \vv_{\vk}| \!\! >=\!0)$
leading effectively to
$
[v^{(i)}_{\vk} \! , v^{(j)}_{\vk'}]
\!=\!
0
$.
He transformed the quantum-fluid Hamiltonian
(\ref{exactH})
to one in configuration space
and obtained the classical-fluid Hamiltonian
for the case of irrotational flow
\cite{SYN2.62,SYN3.62}.
From this view point,
very recently, we have developed
a quantum-fluid approach
to vortex motion in nuclei
\cite{NishProvi3.17}.
In this work we have introduced
Clebsch parameterization
making it possible to describe canonically
quantum-fluid dynamics
and have done Clebsch-parameterized gauge potential
possessing Chern-Simons number
(quantized helicity)
\cite{JackiwPi.00,JackiwNairPi.00,Jackiw.02}.
A study of Chern-Simons form
shed new light on the aspect of 
the behavior of the quantum fluid.

\vspace{0.5cm}


\noindent
\centerline{\bf Acknowledgements}

\vspace{0.3cm}

One of the authors (S.N.) would like to
express his sincere thanks to
Professor Constan\c{c}a Provid\^{e}ncia for
kind and warm hospitality extended to him at
the Centro de F\'\i sica, Universidade de Coimbra.
This work was supported by FCT (Portugal) under the project
CERN/FP/83505/2008.
The authors thank the Yukawa Institute for Theoretical Physics
at Kyoto University. Discussions during the workshop
YITP-W-17-08 on ``Strings and Fields 2017''
are useful to complete this work.


\end{document}